\documentclass[preprint,one-column, double-spaced]{revtex4-2}
\usepackage{amssymb}
\usepackage{amsmath}
\usepackage{graphicx}
\usepackage{textcomp}
\usepackage{titlesec}
\usepackage{color}
\usepackage{setspace}
\begin{document}

\begin{center}
		\vspace{0.5cm}
\huge	Temperature dependence of the lower critical field of the noncentrosymmetric superconductor\,$\alpha$-BiPd

\vspace{1.5cm}

\large	J.\,Juraszek$^{1}$, M.\,Konczykowski$^{2}$, D.\,Kaczorowski$^{1}$, and T.\,Cichorek$^{1}$
\vspace{0.2cm}

\normalsize \textit{$^{1}$Institute of Low Temperature and Structure Research,
	\\
	Polish Academy of Sciences, 50-422 Wroc\l aw, Poland}

\vspace{0.2cm}

\normalsize \textit{$^{2}$Laboratoire des Solides Irradi\'es, CEA/DRF/IRAMIS, \'Ecole Polytechnique,
	\\
	CNRS, Institut Polytechnique de Paris, Palaiseau, F-91128, France}

\end{center}


\vspace{-1cm}
\section{\Large{Abstract}}
Temperature variation of the lower critical field in the noncentrosymmetric superconductor $\alpha$-BiPd was probed by local magnetization measurements using Hall micromagnetometry, performed down to 0.3\,K in a magnetic field applied along the crystallographic $b$ axis.  Below a critical temperature $T_c$\,$\simeq$\,3.8\,K, a conventional $H_{c1}(T)$ dependence was found, typical for a single-band $s$-wave BCS superconductor. The obtained data imply an absence of spin-triplet component in the superconducting wavefunction and marginal multiband effects in this material, which contradicts some literature reports.

\vspace{-1cm}
\section{\Large{Introduction}}

A sufficiently large antisymmetric spin-orbit coupling in superconductors without inversion symmetry  opens the possibility of a novel pairing state which is a mixture of spin-singlet and spin-triplet components. These parity-breaking materials have attracted much theoretical and experimental interest, as they make it a promising route to realize an intrinsic topological superconductivity \cite{SatoRPP2017,SmidmanRPP2017}. Indeed, a singlet-triplet mixing driven by Rashba splitting is expected to give rise to zero-energy Majorana modes, which have potential applications in quantum computing. Moreover, this mixing leads to two-gap physics that significance is determined by the pairing interactions as well as the spin-orbit coupling.  However, evidence of a triplet component in the order parameter of noncentrosymmetric superconductors is still elusive, and presence of multiple conduction and valence bands raises an additional question about multigap superconductivity that makes an insight into this material system even more complex.

The compound $\alpha$-BiPd is believed to be the first material identified as superconducting \cite{Alekseevskii1952} and noncentrosymmetric \cite{Kheiker}. It has a monoclinic crystal structure (space group $P2_1$) of its own type (see critical discussion in Ref. \cite{Yaresko2018}), characterized by very weak chemical bonds between Bi atoms along the $b$ axis \cite{Bhatt1979}. The superconductivity emerges below a critical temperature $T_c$\,$\simeq$\,3.8\,K at ambient conditions and bears a renewed interest due to strong antisymmetric spin-orbit interaction that generates large Rashba splitting. The latter effect was clearly evidenced experimentally by means of angle-resolved photoemission spectroscopy (ARPES) \cite{Neupane2016,Benia2016,Thirupathaiah2016,Lohani2017,Pramanik2021} and scanning tunneling spectroscopy \cite{Benia2016,Sun2015}, and confirmed by ab-initio electronic band structure calculations \cite{Yaresko2018,Neupane2016,Benia2016,Thirupathaiah2016,Lohani2017,Sun2015,Khan2019,Klotz2020}. Furthermore, the spin-resolved ARPES spectra revealed the presence of Dirac surface states \cite{Neupane2016} that were also inferred from the  de Haas - van Alphen oscillations \cite{Khan2019,Klotz2020}. Odd-parity order parameter in $\alpha$-BiPd was hypothesized  from the $^{209}$Bi nuclear quadrupole resonance study, where strongly reduced coherence peak just below $T_c$ in the spin-lattice relaxation rate was found \cite{Matano2013}. Unconventional Cooper paring was also anticipated from the point-contact Andreev reflection spectra \cite{Mondal2012} and the penetration depth data \cite{Jiao2014}. In both experiments, however, the presence of two gaps was assumed with a BCS-type behavior in the $ac$ plane and a power-law temperature variation for the out-of-plane direction. Multiband aspect of superconductivity in $\alpha$-BiPd has been put forward in a few other studies showing, e.g., an upward curvature in the temperature dependence of the upper critical field $H_{c2} (T)$ close to zero-field $T_{c}$ \cite{Mondal2012,Jha2016,Kannan2019}.  Nevertheless, numerous further experiments, including the extensive investigations of $H_{c2}(T)$ down to 30\,mK, have characterized $\alpha$-BiPd as a single-gap nodeless dominantly $s$-wave superconductor \cite{Peets2016}. In addition to this, the order-of-magnitude discrepancy in the reported values of the upper critical field $H_{c2}$ \cite{Sun2015,Mondal2012,Peets2016,Joshi2011,Okawa2013} raises a need for verification of two-band scenario \cite{Jiao2014}. And yet, the recently observed $\pi$ phase shift in oscillations of $T_c$ as a function of the applied magnetic flux threading through a mesoscopic polycrystalline $\alpha$-BiPd rings has been reported as a firm evidence for the spin-triplet pairing state \cite{Xu2020}.

Here, we critically addressed the issue of multiband superconductivity in $\alpha$-BiPd via examining the temperature dependence of the lower critical field $H_{c1}(T)$. The $H_{c1}(T)$ behavior was inspected by means of local magnetization measurements performed down to 0.3\,K. As the experimental technique we used Hall micromagnetomery, which minimizes disadvantages caused by (i) existence of various surface barriers that inhibit penetration of magnetic field, hence leading to overestimation of $H_{c1}$, and (ii) distortion of magnetic field around finite-size sample that brings about underestimation of $H_{c1}$. Our results obtained for single-crystalline $\alpha$-BiPd in magnetic fields applied along the crystallographic $b$ axis indicate a single-gap superconductivity with conventional $s$-wave pairing.

 \vspace{-1cm}
\section{\Large{Results and Discussion}}\label{results}

\vspace{-0.5cm}
\subsection{Electrical Resistivity and Specific Heat}

Figure 1(a) presents the temperature dependence of the electrical resistivity $\rho(T)$ of $\alpha$-BiPd (sample 1) measured with current flowing within the crystallographic $a$-$c$ plane. At 4.2 K, the resistivity is as small as 0.36(4)\,$\mu \Omega$ cm, and increases with increasing temperature up to about 40.5(4.5)\,$\mu\Omega$ cm at 300 K (inset). The residual resistivity ratio RRR\,=\,$\rho(300 K)/\rho(4.2 K)$\,=\,113 is similar to that reported in the literature \cite{Jiao2014,Peets2016,Joshi2011}. Another indication of good quality of our single crystals is a relatively large value of $T_c$\,=\,3.85(2)\,K and a narrow width of the superconducting transition $\Delta T_c$\,$\simeq$\,0.29\,K (estimated from the 10\%–90\% criterion).

The low-temperature dependencies of the specific heat $C(T)$ of $\alpha$-BiPd (sample 2), measured in zero and finite magnetic fields, are depicted in Figure 1(b). A pronounced superconducting anomaly found in $H$\,=\,0 (red points) is significantly depressed in a rather weak external field of 50\,mT (grey points). For 100 mT, we note an absence of superconducting anomaly down to $T$\,=\,0.4\,K, in concert with the upper critical field $\mu_{0}$$H_{c2}(0)$ $\simeq$ 89 mT along the [010] direction reported in Ref. \cite{Peets2016}. From these normal-state data, we estimated the electronic  $C_e$\,=\,$\gamma$$T$ and phonon $C_{ph}$ contributions, where $\gamma$\,=\,3.9(2)\,mJmol$^{-1}$K$^{-2}$ is the Sommerfeld coefficient. The inset of Figure 1(b) presents the electronic specific heat of $\alpha$-BiPd in the vicinity of $T_c$, as $C_e$/T vs temperature. Using an equal-entropy construction, one finds for this crystal $T_c$\,=\,3.70(5)\,K, $\Delta T_c$\,$\simeq$\,0.09\,K, and a specific heat jump $\Delta C_e$/$\gamma T_c$\,$\simeq$\,1.72. The superconducting discontinuity is somewhat larger than that predicted by the BCS theory (=\,1.426), but very close to that reported in Ref. \cite{Sun2015}. On the contrary, it notably differs from the reduced specific heat jump given in Ref. \cite{Joshi2011}, where it was discussed in the context of a multiband superconductivity. It is also worth noting that the gap size $\Delta$(0) = 1.94$k_{\rm{B}}T_c$ found for our crystal is much larger than $\Delta$(0) = 1.35$k_{B}T_c$ derived from the $^{209}$Bi nuclear quadrupole resonance measurements \cite{Matano2013}. In the latter study, the suppressed and broad coherence peak below $T_c$ in the spin-lattice relaxation rate was suggested to be a signature of spin-triplet component in the superconducting state of $\alpha$-BiPd.

\vspace{-0.5cm}
\subsection{Magnetization Measurements}
For three other $\alpha$-BiPd samples used in our $H_{c1}$ study, we have determined their temperature dependencies of the magnetization $M(T)$ in the vicinity of superconducting transition at $\mu_0$$H$\,=\,1\,mT, as depicted in the insets of Figures 2(a-c). Whereas a drop of $M(T)$ signals the onset of a superconducting phase transition at virtually the same $T_c$$^{onset}$ of 3.8\,K, a somewhat sharper transition $\Delta T_c$\,=\,0.05(1)\,K was found for sample B with $T_c$\,=\,3.77(2)\,K. For sample C, we have measured both the zero-field cooled (ZFC) and field-cooled (FC) magnetization, as shown in the lower inset of Figure 2(c). When we assume the ZFC data to represent the complete diamagnetic shielding below 3.6\,K, we can conclude from the FC results that about 36\% volume of the $\alpha$-BiPd sample was in the Meissner-Ochsenfeld state at $\mu_{0}H$\,=\,1\,mT below this temperature. The main panels of Figures 2(a) and 2(b) present the initial parts of the magnetization curves of two $\alpha$-BiPd samples measured using a Hall micromagnetometry at different temperatures down to 0.3\,K\,($\simeq$\,0.08$T_c$). For comparison, in Figure 2(c) there are shown the $M(H)$ curves measured for sample C using a commercial SQUID magnetometer. These bulk magnetization measurements were performed down to 2 K\,($\simeq$\,0.53$T_c$) only.

In Figures 3(a) and 3(b) we show the local magnetization $\Delta M(H)$ curves obtained at different temperatures after subtraction of the Meissner (shielding) slope for samples A and B, respectively. (See Figure S1 for similar results for sample C.) When an applied  field reaches a critical value, called the field of first flux penetration $H_{\rm{p}}$, the first Abrikosov vortex enters the sample and more vortices penetrate it with the further increasing magnetic field. Thus, $H_{\rm{p}}$ corresponds to the field at which a rapid rise of $\Delta M(H)$ sets in, as indicated by black triangles. An $H_{c1}$ value  at given temperature can be directly estimated from the geometric conversion factor. According to Ref. \cite{Joshi2019}, for a cuboid with dimensions $2m \times 2n \times 2k$ placed in a magnetic field oriented along the $k$ direction, the appropriate approximation of $H_{c1}$ is given 
by
\begin{equation}
	\label{eqVline}
	H_{c1}=\frac{H_{\rm{p}}}{1+\chi N},
\end{equation}

where $\chi$ is the dimensionless intrinsic magnetic susceptibility of the material in the superconducting state which can be taken to be equal to $-$1, and $N$\,=\,[$ 1+\frac{3}{4}\frac{k}{m}\left ( 1+\frac{m}{n}\right )]^{-1}$ stands for the effective demagnetization factor (see Supporting Table 1). In our experiments, 2$m$\,=\,40\,$\mu$m and 2$n$ = 40 $\mu$m correspond to the width and length of the Hall sensor active area, respectively, and 2$k$ is the thickness of the measured $\alpha$-BiPd specimen along the crystallographic $b$ axis.

In the Ginzburg-Landau theory, the lower critical field is given by 
\begin{equation}
	\label{Hc1a}
	H_{c1}(T)\,=\,\frac{\phi _{0}}{4\pi \mu _{0}\lambda ^{2}(T)}\left ( \rm{ln}\kappa(\textit{T}) + 0.5 \right ),
\end{equation}
where $\kappa$($T$)\,=\,$\lambda$($T$)/$\xi$($T$) is the Ginzburg-Landau parameter and $\xi$($T$) is the coherence length. Since the dimensionless parameter $\kappa(T)$ is effectively constant under the logarithm, the behavior of $H_{c1}$$(T)$\,$\propto$\,$\lambda ^{-2}(T)$ provides a powerful way to deduce the temperature dependence of the superfluid density, and hence the symmetry of Cooper pairing as well as  the presence of multiband effects.

Figure\,4 displays the $H_{c1}(T)$ dependencies derived for $\alpha$-BiPd in $H$\,$\parallel$\,[010] using Eq.\,1 and the $\Delta M(H)$ data shown in Figure\,3. At $T$\,=\,0.3\,K, the values of $\mu_{0}H_{c1}$ are equal to 20.9(2) and 20.5(2)\,mT for samples A and B, respectively. Small difference between these values likely arises from uncertainty in the geometrical factor, and their average $\mu_{0}H_{c1}$ = 20.7(2)\,mT is in good agreement with the lower limit of the literature data, which span from 20 to 28\,mT \cite{Sun2015,Jha2016,Kannan2019}. Relatively large spread of about of 40\% in the hitherto reported vales of $\mu_{0}H_{c1}$(0) can be attributed mainly to unspecified geometric conversion factors and some ambiguity in the determination of $H_{\rm{p}}$. Besides, it should be noted that the majority of the previous estimations of $H_{c1}$(0) was based on experiments performed at relatively high temperatures $T$\,$\geq$\,2\,K, i.e., above about 0.5$T_c$.

The $H_{c1}(T)$ dependencies for $H$\,$\parallel$\,[010] presented in Figure\,4 point at a clear picture of $\alpha$-BiPd being a single-gap $s$-wave superconductor with a somewhat enhanced electron-phonon coupling. Following the elaboration of the alpha-model derived from the BCS theory of superconductivity \cite{Johnston2013}, this is illustrated by the solid lines calculated with the superconducting gap $\Delta(0)$=2.0$k_B T_c$ which is slightly larger that that given by the conventional BCS relation $\Delta(0)$=1.764$k_B T_c$. Most importantly, the $H_{c1}(T)$ curve lack a singularity down to at least 0.08$T_c$ that would hint at multiband effects \cite{Juraszek}. The single-band parabolic behavior of the lower critical field in $\alpha$-BiPd is additionally highlighted in the inset of Figure\,4, where the reduced $H_{c1}$ values are plotted as a function of ($T/T_c$)$^2$.

The comprehensive investigation of the upper critical field of $\alpha$-BiPd revealed $\mu_{0}H_{c2}$(0)\,$\simeq$\,75\,mT for $H$\,$\parallel$\,[010]  \cite{Peets2016}, which is by more than an order of magnitude smaller than the values reported earlier  \cite{Mondal2012,Joshi2011,Okawa2013}. This zero-temperature value of the upper critical field combined with our estimation of $\mu_{0}H_{c1}$(0)\,$\simeq$\,20.7\,mT yields the out-of-plane thermodynamic critical field $\mu_{0}H_c$\,$\simeq$\,($H_{c1} H_{c2}$)$^{1/2}$\,$\simeq$\,39.4\,mT that is in good agreement with that determined from the specific-heat measurements \cite{Peets2016}. Remarkably, it implies a very narrow interval for an external field penetrating the sample as non-overlapping vortices. In other words, the relation 2$H_c \simeq H_{c2}$ found for $H$\,$\parallel$\,[010] calls for reinterpretation of the field-dependent experiments, which were argued to provide the evidence for multiple superconducting energy gaps in $\alpha$-BiPd \cite{Mondal2012,Jiao2014,Yan2016}.  

\vspace{2mm}

Contrary to common belief, a large number of bands crossing the Fermi level (approximately 13 in the case of $\alpha$-BiPd \cite{Neupane2016,Sun2015,Klotz2020,Peets2016}) does not necessarily promote multiband superconductivity. This is because a variety of interband interactions tend to smear out specific characteristics of singular gaps, and thus give rise to an effective single-gap behaviour \cite{Zehetmayer2013}. Therefore, in view of the $H_{c1}(T)$ results  shown in Figure 4, the preceding observations of the upward curvature of $H_{c2}(T)$  for $\alpha$-BiPd \cite{Mondal2012,Jha2016,Kannan2019,Peets2016} may reflect other effects caused by extrinsic mechanisms. In particular, it was theoretically proposed that such an unusual  $H_{c2}(T)$ behavior can be a consequence of spatial dependence of the coherence length associated with a modulation of disorder characteristics, such as the diffusion coefficient or the mean-free path, that frequently happens in superconducting layered structures due to, e.g., fluctuations in the crystal axes orientation \cite{Kopasov2017,Antonov2020}. The latter scenario seems probable to occur in $\alpha$-BiPd that exhibits the layered crystal structure with an easy-cleavage (010) plane which favors an occurrence of step edges on the surface \cite{Sun2015}, and thus a locally different field distribution can mimic fluctuations in the crystal axes orientation.

The conventional behavior of $H_{c1}(T)$ along the $b$ axis established in our study does not support the scenario of spin-triplet component in the superfluid density of bulk $\alpha$-BiPd samples suggested in the literature \cite{Mondal2012,Jiao2014,Xu2020}. In particular, the well-defined single-band $s$-wave picture is in striking contrast to  the results of the Little-Parks experiment on mesoscopic polycrystalline $\alpha$-BiPd rings which have been ascribed to  half-integer magnetic flux quantization, an expected feature of spin-triplet pairing \cite{Xu2020}. Furthermore, the fact that the period of Little-Parks oscillations remains unchanged in fields much higher than those corresponding to the $H_{c2}$ values for bulk samples raises doubts about a mixture of singlet-triplet pairing in $\alpha$-BiPd. It should be mentioned, however, that the observation of half-quantum flux was reported before for mesoscopic rings made of another Pd-Bi superconductor $\beta$-Bi$_2$Pd with $\mu_{0}H_{c2}$\,$\simeq$\,0.7\,T \cite{Li2019}. Also for this compound, the symmetry of the superconducting order parameter remains unsolved. While Hall-probe magnetometry and scanning tunneling spectroscopy strongly supported single-gap $s$-wave superconductivity in bulk specimens \cite{Kacmarcik2016}, other tunneling spectroscopy studies reported realization of Majorana-bound states at the center of the vortices in epitaxial thin films \cite{Lv2017}.

\vspace{-1cm}
\section{\Large{Conclusion}}
In conclusion, we have determined the temperature dependence of the lower critical field of $\alpha$-BiPd down to $\sim$0.08$\textsl{T}_c$ along the monoclinic $b$ axis. The $H_{c1}(T)$ results indicate that this noncentrosymmetric compound should be considered a single-gap $s$-wave superconductor. We found no signature of an unconventional contribution to the superconducting gap. We also did not observe any hint at multiple superconducting gaps. The zero-temperature value of $H_{c1}$ being by only a factor of $\sim$3.7 smaller than $H_{c2}(0)$ points out the need for a reinterpretation of the field-dependent thermodynamic data hitherto reported for $\alpha$-BiPd, which have been used as an argument for the claim of multiple gaps. In particular, we suggest that the observed upward curvature of $H_{c2}(T)$ in the vicinity of $T_{c}$ reflects an extrinsic effect closely associated with a layered crystal structure of the compound.

\vspace{-1cm}
\section{\Large{Experimental Section}}\label{experiment}
Single crystals of $\alpha$-BiPd were grown by a modified Bridgman method, following a recipe described in the literature \cite{Joshi2011}. Powder and single-crystal X-ray diffraction as  well as a back-reflection Laue method (see Supporting Figure 1) were used to prove their noncentrosymmetric monoclinic crystal structure (space group $P2_1$), and energy dispersive X-ray spectroscopy was applied to verify their stoichiometric, highly homogeneous chemical composition. Further quality check of the $\alpha$-BiPd crystals was made by means of electrical resistivity and heat capacity measurements performed using a Quantum Design PPMS-9 platform. Inter alia, the results revealed for each single-crystalline specimen a sharp superconducting transition at $T_c$\,$\simeq$\,3.75 K, in agreement with the other works \cite{Matano2013,Mondal2012,Jiao2014,Jha2016,Kannan2019,Peets2016} . The specific heat was measured in the temperature range between 0.4 and 4 K in magnetic fields up to 100\,mT. The zero-field electrical resistivity was measured in the temperature interval 2--300 K using a standard ac four-probe technique and electric leads attached to a bar-shaped specimen with silver epoxy paste.

For low-temperature magnetization measurements, we used three single crystals grown in the same batch. Samples A and B, were investigated using a Hall micromagnetometry.  In our local magnetization experiments performed on a $^{3}$He cryostat operating down to 0.3\,K, each sensor has the active area of 40 $\mu$m $\times$ 40 $\mu$m and the distance between a sample and a Hall sensor amounts to 1500 \AA. A field-sweep rate as small as 50\,$\mu$T/min was applied to ensure a high accuracy. After each sweep up to 20\,mT, the sample was warmed above $T_c$ and cooled again in zero applied field. An identical sensor was used to monitor an applied field $H$ that was oriented perpendicular to the $a$-$c$ plane of the crystal. More details about our Hall-micromagnetometry setup can be found elsewhere\,\cite{Juraszek}. For the magnetization measurement on sample C, we used a Quantum Design superconducting quantum interference device magnetometer (MPMS-XL) operating on a $^{4}$He cryostat. Unfortunately, a brittleness of crystals and a layered structure of the compound make challenging the preparation of small and thin samples suitable for the  magnetization measurements in the $a$-$c$ plane.

\textbf{Acknowledgements} \par 
D.K. thanks the National Science Centre (Poland) for the Research Grant 2021/40/Q/ST5/00066. 
J.J. was supported by the Foundation for Polish Science (FNP), program  START  036.2021.

	\vspace*{2.4cm}

\newpage

\begin{figure*}[h]
	\centering
	\includegraphics[width=125mm,keepaspectratio=true]{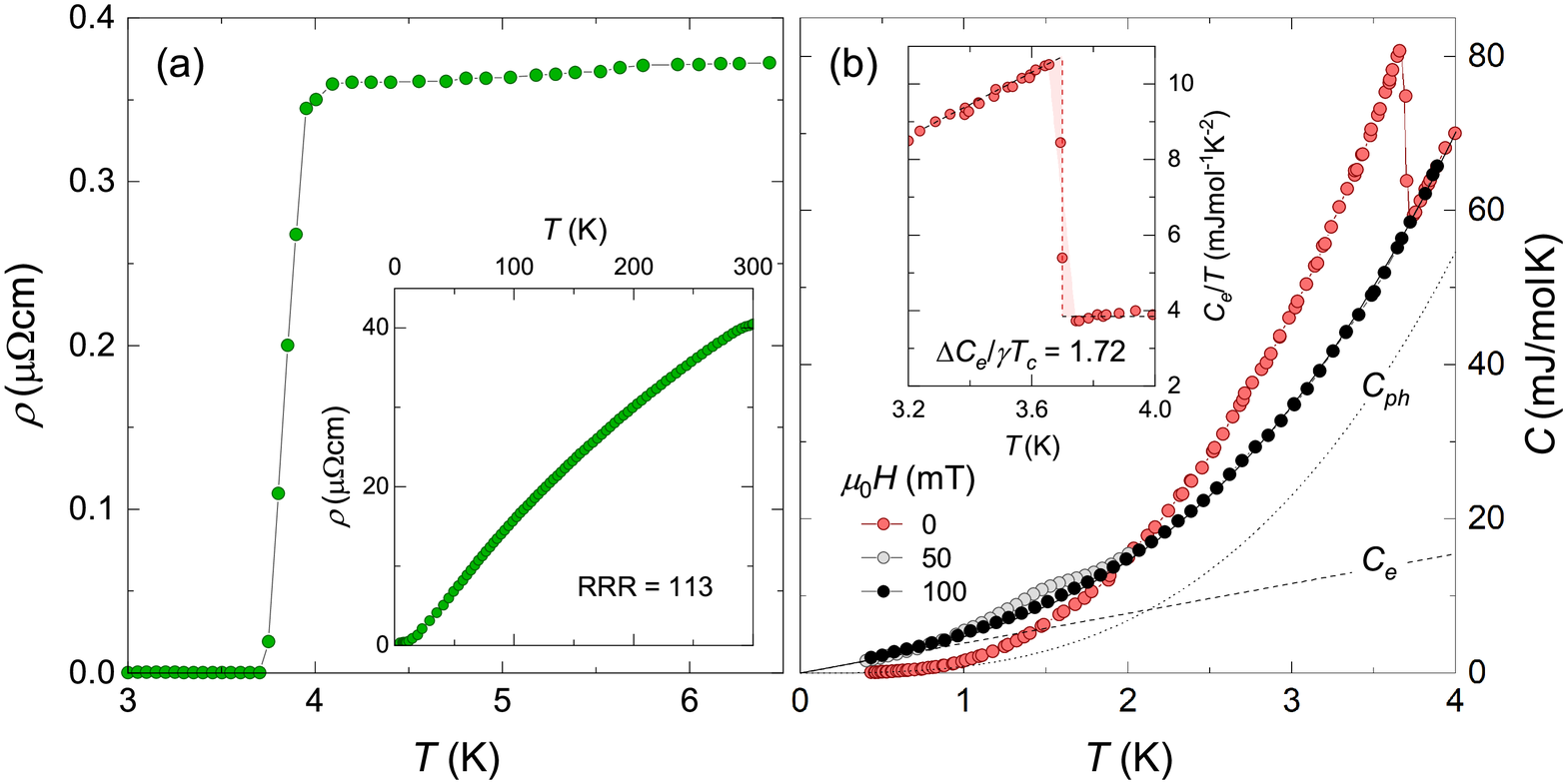}
	\caption{\textbf{Characteristics of the $\alpha$-BiPd single crystals.} (a) Low-temperature electrical resistivity in the $a$-$c$ plane showing a sudden drop to zero at $T_c$\,=\,3.85(2)\,K (sample 1). Inset: The $\rho$($T$) dependence in the temperature range 2\,--\,300\,K. (b) Low-temperature specific heat taken in zero and finite magnetic fields applied perpendicular to the $a$-$c$ plane (sample 2). The dashed and dotted lines represent the electronic $C_e$ and lattice $C_{ph}$ contributions, respectively, in the 100 mT data. Inset: The ratio $C_e$/$T$ versus $T$ in the vicinity of $T_c$\,=\,3.70(5)\,K  and an equal-entropy construction.}
	\label{FIG1}
\end{figure*}

\begin{figure*}[h]
	\centering
	\includegraphics[width=155mm,keepaspectratio=true]{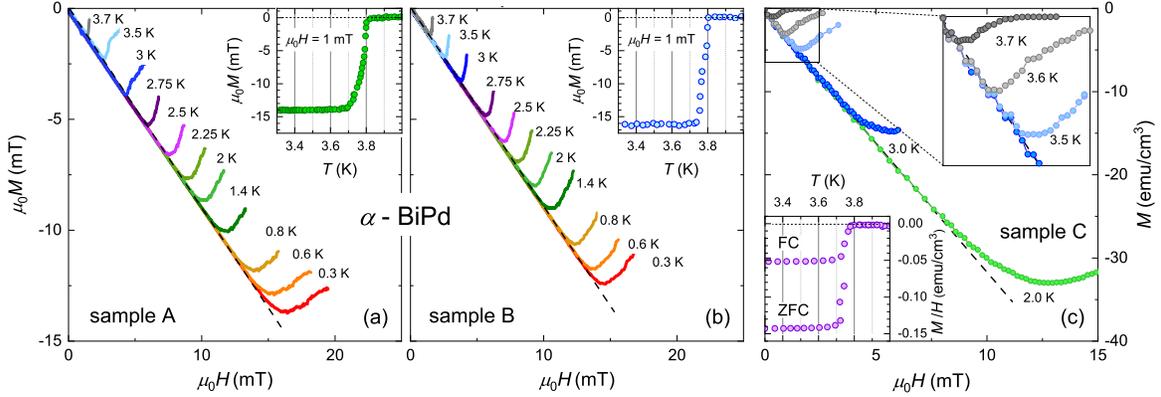}
	\caption{\textbf{Low-temperature magnetization of $\alpha$-BiPd for $H$\,$\parallel$\,[010].} (a) Local magnetization isotherms taken at different temperatures down to 0.3\,K for sample A. The dashed line marks a shielding slope, which is slightly smaller from unity due to a flux leakage around the sample (cf. Experimental Section). Inset: The ZFC magnetization measured in $\mu_{0}H$\,=\,1\,mT. (b) Similar $M(H)$ and $M(T)$ results for sample B. (c) The bulk $M(H)$ data for sample C obtained down to 2\,K. Upper inset: Zoom into the $M(H)$ curves measured at temperatures close to zero-field $T_c$. Lower inset: The ZFC and FC curves taken in $\mu_{0}H$\,=\,1\,mT.}
	\label{FIG2}
\end{figure*}

\begin{figure}[h]
	\centering
	\includegraphics[width=0.5\textwidth]{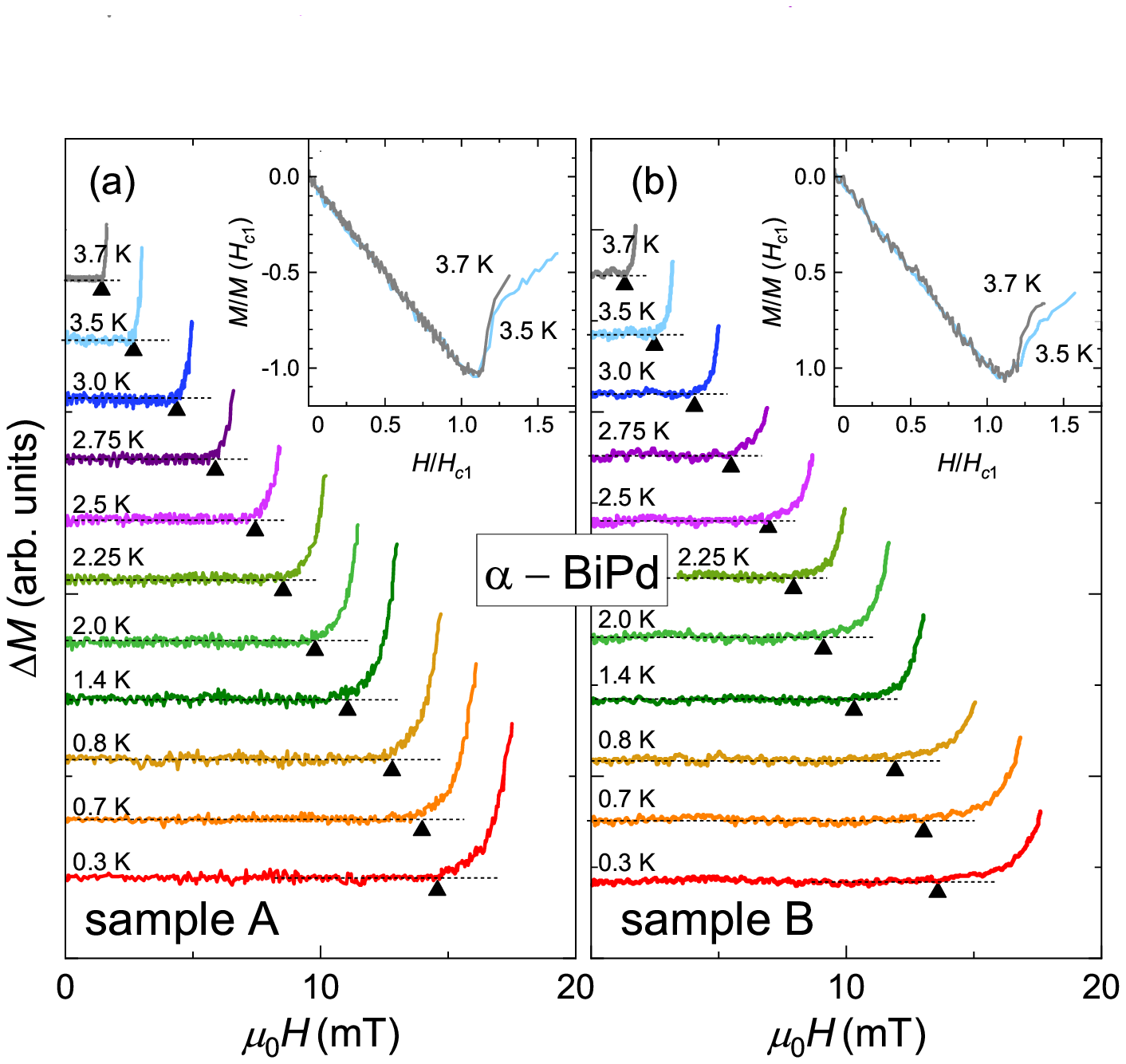}
	\caption{\textbf{Field of first flux penetration in $\alpha$-BiPd for $H$$\,\parallel$\,[010].} (a) Local magnetization in sample A obtained after subtraction of the Meissner slope from the $M(H)$ data shown in Figure 2(a).  For clarity, each isotherm was shifted vertically. The triangles indicate the field of first flux penetration $H_{\rm{p}}$. Inset: The normalized $M(H)$ dependencies measured at temperatures close to zero-field $T_c$. (b) Similar results obtained for sample B.}
	\label{FIG3}
\end{figure}

\begin{figure}[h]
	\centering
	\includegraphics[width=0.5\textwidth]{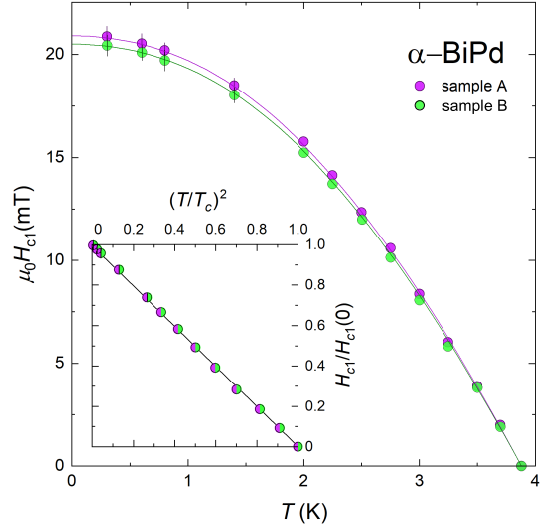}
	\caption{\textbf{The lower critical field of $\alpha$-BiPd along the $b$ axis.} (a) Temperature dependencies of $H_{c1}$ determined for samples A and B. The solid lines represent $H_{c1}(T)$ calculated with  $\Delta(0)$=2.0$k_B T_c$. Above 2 K, error bars are the same size or smaller than the symbols. Inset: Normalized lower critical field as a function of the square of the normalized temperature.}
	\label{FIG4}
\end{figure}

\clearpage
\vfill
\newpage
\begin{center}
	\vspace*{2.4cm}	
	\huge Supporting Information	
	\vspace{0.5cm}
\end{center}

\renewcommand{\figurename}{Supporting Figure}
\setcounter{figure}{0}
\vfill
\vspace{-2cm}
\begin{figure*}[h]
\centering
\includegraphics[width=80mm,keepaspectratio=true]{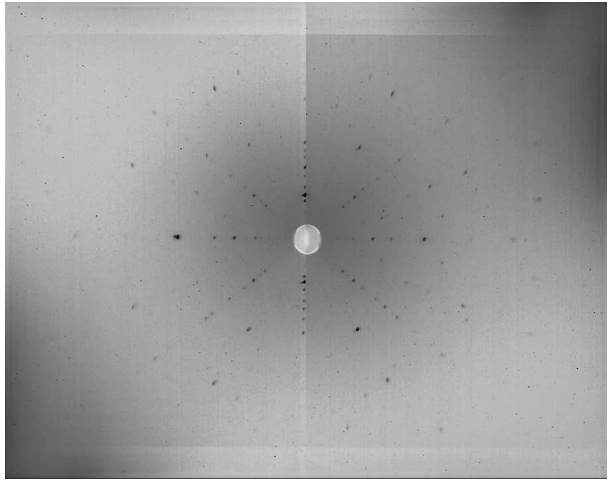}
\caption{\textbf{Back-reflection Laue photographs of the $\alpha$-BiPd single crystal}. Figure shows the diffractogram along the monoclinic $b$ axis.}
\label{FIG2}
\end{figure*}

\begin{table}[h]
\caption{\label{tab:table} Dimensions of three $\alpha$-BiPd samples used in this study for $H$\,$\parallel$\,[010]: $m$ and $n$ are the sample length and width in the \textit{a-c} plane and $k$ is the dimension along the $b$ axis. For sample A and B, $m$ and $n$ correspond to the width and length of the Hall sensor active area. $N$ is the effective demagnetization factor  and  $H_{c1}$(0) is the lower critical field in the limit $T$\,=\,0.}
\vspace{2mm}
\begin{tabular}{ccccccccc}
	\hline  \hline
	Sample   & 2$m$  [mm] & 2$n$  [mm] & 2$k$  [mm]   &  $N$ & $\mu_{0}H_{c1}$(0)  [mT]  \\ \hline
	A      &  0.04 & 0.04  &    0.06  &  0.31  & 20.9  \\
	B      &  0.04 & 0.04  &   0.05  &  0.34 & 20.7   \\
	C      & 0.76  & 0.97  &    0.15   &  0.79* & 48.4  \\ 
	\hline  \hline
\end{tabular}
\end{table}
\noindent*Note that $N$\,=\,0.74 for sample C can be alternatively estimated from the slope of the initial magnetization curves shown in Fig. 2(c).

\vspace{10cm}
\begin{figure*}
\centering
\includegraphics[width=55mm,keepaspectratio=true]{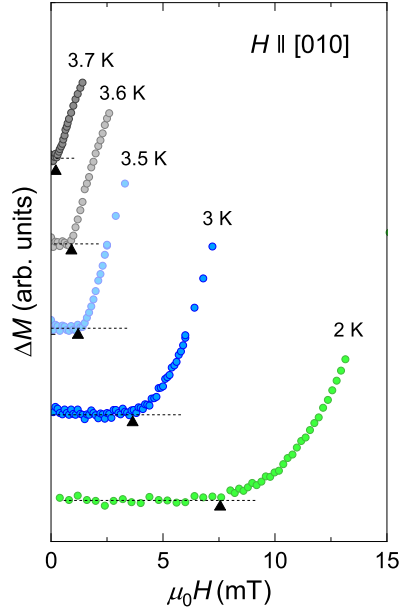}
\caption{\textbf{Field of first flux penetration in sample C of $\alpha$-BiPd for $H$$\,\parallel$\,[010]}. Bulk magnetization in sample C obtained after subtraction of the Meissner slope from the $M(H)$ data shown in Figure 2(c).  For clarity, each isotherm was shifted vertically. The triangles indicate the field of first flux penetration $H_{\rm{p}}$.} 
\label{FIG2}
\end{figure*}

\vfill
\vspace{10cm}

\newpage
\vspace{10cm}
\vfill

\begin{figure*}[htb]
\begin{center}
\includegraphics[width=90mm,keepaspectratio=true]{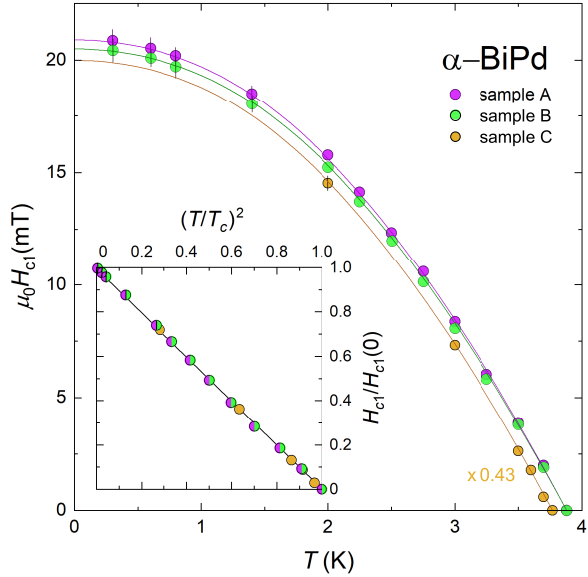}
\caption{\textbf{The temperature dependence of $H_{c1}$ of $\alpha$-BiPd for $H$\,$\parallel$\,[010].} For all samples investigated, the $H_{c1}$($T$) data  follow the single-band $s$-wave dependence calculated with  $\Delta(0)$=2.0$k_B T_c$, as shown by the solid lines. Note larger values of $H_{c1}$ for sample C compared to those estimated using a Hall micromagnetometry.  Inset: The $H_{c1}$/$H_{c1}(0)$ data versus the square of the normalized temperature.}
\label{fig:FigS2}
\end{center}
\end{figure*}

\end{document}